\documentclass[fontsize=10pt, DIV=calc, a4paper, twocolumn, headings=standardclasses]{scrartcl}

\usepackage[utf8]{inputenc}
\usepackage{lmodern}
\usepackage{placeins}

\usepackage[english]{babel}
\usepackage[activate={true, nocompatibility}, final, tracking=true, kerning=true, factor=1100, stretch=10, shrink=10]{microtype}
\usepackage{booktabs, graphicx, amsmath, amssymb, physics, textcomp}
\usepackage{newtxtext,newtxmath}
\usepackage[format=plain, labelfont=bf]{caption}
\usepackage[auth-lg, affil-it]{authblk}
\usepackage[compress, numbers, sort]{natbib}
\usepackage[hidelinks]{hyperref}

\title{Simulating Organogenesis in COMSOL Multiphysics\textsuperscript{\textregistered}: \\ Tissue Patterning with Directed Cell Migration}

\author[$1,2$]{Malte Mederacke}
\author[$1$]{Chengyou Yu}
\author[$1,2$]{Roman Vetter}
\author[$1,2$]{Dagmar Iber}
\affil[$1$]{Department of Biosystems Science and Engineering, ETH Z\"{u}rich, Schanzenstrasse 44, 4056 Basel, Switzerland}
\affil[$2$]{Swiss Institute of Bioinformatics, Schanzenstrasse 44, 4056 Basel, Switzerland}
\date{}

\begin{document}

\maketitle

\noindent\textbf{Abstract} \\
We present a COMSOL Multiphysics\textsuperscript{\textregistered} implementation of a continuum model for directed cell migration, a key mechanism underlying tissue self-organization and morphogenesis. The model is formulated as a partial integro-differential equation (PIDE), combining random motility with non-local, density-dependent guidance cues to capture phenomena such as cell sorting and aggregation. Our framework supports simulations in one, two, and three dimensions, with both zero-flux and periodic boundary conditions, and can be reformulated in a Lagrangian setting to efficiently handle tissue growth and domain deformation. We demonstrate that COMSOL Multiphysics\textsuperscript{\textregistered} enables a flexible and accessible implementation of PIDEs, providing a generalizable platform for studying collective cell behavior and pattern formation in complex biological contexts. \\
\vspace{1em}


\vspace{1.5em}

\section*{Introduction}

In this article, we present our implementation of a continuum model for Directed Cell Migration (DCM) \cite{yu2025directed}, a fundamental biological mechanism by which cells self-organize into complex tissues during development, in COMSOL Multiphysics\textsuperscript{\textregistered}. We have previously made extensive use of COMSOL Multiphysics\textsuperscript{\textregistered} to simulate biological self-organization through reaction–diffusion mechanisms \cite{Germann:2011, Menshykau:2012, Menshykau:2013, Vollmer:2013, Karimaddini:2014, Wittwer:2016, Wittwer:2017}, tissue mechanics \cite{Peters:2017}, and fluid flow \cite{Conrad:2021}. We implemented these models on growing, deforming, or interacting domains using Arbitrary Lagrangian--Eulerian (ALE) \cite{Menshykau:2012} or phase-field methods \cite{Wittwer:2016, Wittwer:2017}, and represented individual cells with compartmentalized diffusion and signaling \cite{Vollmer:2013, Kurics:2014}. Moreover, we implemented efficient methods for parameter estimation \cite{Menshykau:2013}, and to use image-derived geometries \cite{Karimaddini:2014}.

DCM enables the formation of spatial patterns from initially homogeneous conditions, relying solely on interactions on the cellular scale, to pattern tissue on the much larger organ scale. DCM arises from the interplay between random cellular motility and guided movement, ultimately leading to phenomena such as cell sorting and aggregation. In biological systems, guidance cues for directed migration include gradients in chemical signals (chemotaxis), substrate stiffness (durotaxis), surface topography (topotaxis), and differences in cell-cell adhesion (differential adhesion), among others.

Numerical modeling is particularly valuable in studying these mechanisms, given the complexity of cell behavior and tissue geometry, which often defy analytical treatment. Continuum models are well-suited to capture tissue-scale dynamics while maintaining a connection to analytical insights. To this end, single cell properties are locally averaged and the population of cells is represented by a cell density field whose dynamics are driven by cellular flow. In our approach, DCM is described using partial integro-differential equations (PIDEs), which are capable of incorporating both local diffusion and non-local interaction terms, essential for modeling realistic cell dynamics. The system employed in this work builds on formulations previously developed \cite{carrillo2019population, falco2023local, armstrong2006continuum}.

Solving PIDEs poses significant numerical challenges and typically requires custom-built solvers based on finite volume or finite element methods \cite{carrillo2019population, falco2023local, armstrong2006continuum, carrillo2025new}. These implementations often lack flexibility and demand expert-level knowledge to modify or extend. To address this limitation, we implement the model in COMSOL Multiphysics\textsuperscript{\textregistered}, which provides the necessary discrete operators to include integral terms in the PDEs. Our approach supports simulations in one, two, and three spatial dimensions and accommodates both zero-flux (Neumann) and periodic boundary conditions, broadening its applicability to various biological scenarios.

Furthermore, tissue growth is an inherent aspect of developmental biology and morphogenesis. To handle domain expansion efficiently, we demonstrate how the DCM model can be reformulated in a Lagrangian framework, enhancing computational performance while maintaining model fidelity.

In summary, we present a generalizable and accessible numerical framework to study a wide range of directed cell migration phenomena, offering new opportunities for exploring pattern formation and organogenesis through simulation.

\section*{Theory}

In this section, we introduce the governing equations of the PIDE-based continuum model for directed cell migration. The model describes the temporal evolution of a cell density field $c(\vec{x},t)$:
\begin{equation}
\frac{\partial c}{\partial t} = -\vec{\nabla}\cdot\vec{j},
\end{equation}
where $\vec{j}$ denotes the cellular flux. This flux can arise from a combination of random or pressure-driven motility and directed, density-dependent migration. For a single-population model, this leads to the following equation in $n$ spatial dimensions: 
\begin{equation}
\vec{j} = - d(c) D \vec{\nabla} c + a h(c) \vec{I}(c,\vec{x}), 
\end{equation}
where
\begin{equation}
\vec{I}(c,\vec{x}) = \int_0^R \int_{S^{n-1}} g\big(c(\vec{x}+r \vec{\eta})\big) r^{n-1} \vec{\eta} \, d\vec{\eta} \, dr.
\end{equation}
Here, $D$ denotes the cellular diffusion coefficient and $a$ the strength of the directed motility forces. 

The model is designed to capture four key aspects of collective cell behavior. First, cells exhibit motility, which can be random (diffusion-like) or pressure-driven, thereby contributing to tissue fluidity. Second, cells are able to sense their surroundings. This is represented by the integral $\vec{I}(c,\vec{x})$, which integrates local densities within a sensing radius $R$, corresponding to a line in 1D, a disk in 2D, or a ball in 3D. Third, cells interact and attract one another with strength $a$ within that radius. For simplicity, this attraction is assumed to be spatially uniform here, i.e., independent of distance within the sensing radius. Finally, the domain has a finite carrying capacity, such that cells locally compete for space. This crowding effect gives rise to a population pressure that counteracts aggregation. In the model, this population pressure can be incorporated either into the random motility term $d(c)$, the directed motility term $h(c)$, or the sensing kernel $g(c)$. In this technical paper, we focus on the special case $d(c) = 1$, $h(c) = c(1-c)$, and $g(c) = 1$. For an exploration of alternative formulations, see \cite{yu2025directed}.

\subsection*{Non-dimensionalization}

For computational convenience we non-dimensionalize the model by setting
\begin{equation}
\vec{\xi} = \frac{\vec{x}}{L_\mathrm{c}}, \quad 
\tau = \frac{Dt}{L_\mathrm{c}^2}, \quad 
\alpha = \frac{a L_\mathrm{c}^{\,n+1}}{D}, \quad 
\rho = \frac{R}{L_\mathrm{c}},
\end{equation}
where $L_\mathrm{c}$ is a characteristic length scale. Here, we choose $L_\mathrm{c} = R$, i.e., $\rho = 1$. The non-dimensionalized equation then reads
\begin{equation}
\label{eq:nd}
\frac{\partial c}{\partial \tau} = \vec{\nabla} \cdot \left(d(c) \vec{\nabla} c\right) - \alpha \vec{\nabla} \cdot \left(h(c) \vec{I}(c,\vec{\xi})\right)
\end{equation}
where
\begin{equation}\label{eq:I_nd}
\vec{I}(c,\vec{\xi}) = \int_0^\rho \int_{S^{n-1}} g\big(c(\vec{\xi}+r \vec{\eta})\big) r^{n-1} \vec{\eta} \, d\vec{\eta} \, dr.
\end{equation}

\subsection*{Domain Growth}

To avoid computationally expensive remeshing during domain growth, we transform the coordinate system from an Eulerian to a Lagrangian frame. This enables the simulation of domain growth on a static geometry. We begin this transformation with the general form of Eqs.~\ref{eq:nd} and \ref{eq:I_nd}, incorporating advection by a velocity field $\vec{u} = \vec{\xi}'(\tau)$ induced by the expansion of the tissue:
\begin{equation}
\frac{\partial c}{\partial \tau} + \vec{\nabla} \cdot (c \vec{u}) = \vec{\nabla} \cdot \left( d(c)\vec{\nabla} c \right) - \alpha \vec{\nabla} \cdot \left( h(c)\vec{I}(c) \right)
\label{eq:nd_velocity}
\end{equation}
To transform Eq.~\ref{eq:nd_velocity} from the Eulerian into the Lagrangian frame, we define the rescaled coordinate
\begin{equation}
\vec{\Xi} = \theta(\tau) \vec{\xi}(\tau)
\end{equation}
where $\theta(\tau) = \partial \Xi_i/\partial \xi_i = L_0/L(\tau)$ is the stretching factor. We consider here uniform linear domain growth of the form
\begin{equation}
L(\tau) = L_0 + v\tau.
\end{equation}
The velocity field $\vec{u}$ can be expressed as:
\begin{equation}
\vec{u} = \vec{\xi}'(\tau) = \vec{\Xi} \frac{L'(\tau)}{L_0}.
\end{equation}
The gradient operator transforms as
\begin{equation}
\vec{\nabla} = \theta(\tau) \vec{\nabla}_\mathrm{L}
\end{equation}
where $\vec{\nabla}_\mathrm{L}$ denotes the gradient in Lagrangian coordinates, i.e., with respect to $\vec{\Xi}$. Since $\vec{u} \cdot \vec{\nabla} c$ vanishes in the Lagrangian frame, and the divergence of the velocity becomes $\vec{\nabla} \cdot \vec{u} = nL'(\tau)/L(\tau)$ for uniform growth in $n$ dimensions, with $L'(\tau) = v$, Eq.~\ref{eq:nd_velocity} transforms into
\begin{equation}
\label{eq:ndL}
\begin{aligned}
\frac{\partial c}{\partial \tau} &= \theta^2(\tau) \vec{\nabla}_\mathrm{L} \cdot \left( d(c) \vec{\nabla}_\mathrm{L} c \right) 
- \theta(\tau) \alpha \vec{\nabla}_\mathrm{L} \cdot \left( h(c) \vec{I}_\mathrm{L}(c, \tau) \right)\\
&- \frac{nvc}{L_0+v\tau}.
\end{aligned}
\end{equation}
The non-local interaction term $\vec{I}_\mathrm{L}$ in Lagrangian then reads
\begin{equation}
\vec{I}_\mathrm{L}(c, \vec{\Xi}, \tau) = 
\int_0^{\theta(\tau)\rho} \int_{S^{n-1}} 
g\big(c(\vec{\Xi} + r \vec{\eta})\big)\, r^{n-1} \vec{\eta}\, d\vec{\eta}\, dr.
\end{equation}

\clearpage\section*{Implementation in COMSOL Multiphysics\textsuperscript{\textregistered}}

All simulations were performed in COMSOL Multiphysics\textsuperscript{\textregistered} v6.3. Eq.~\ref{eq:nd} or \ref{eq:ndL} was implemented as a Coefficient Form PDE.

\subsection*{Implementation of the integral}

In COMSOL Multiphysics\textsuperscript{\textregistered}, depending on the dimensionality of the problem, different operators are available to implement the integral. In 1D, it is relatively straightforward to use the \texttt{intop} function together with the smoothed Heaviside function \texttt{flc2hs} to perform the directed integration. The functions \texttt{diskint(r,expr,m)} in 2D and \texttt{ballint(r,expr,m)} in 3D compute the spherical integral of an expression \texttt{expr} over a neighborhood of radius \texttt{r}. \texttt{m} specifies the number of quadrature points used in the numerical integration. In our case, \texttt{r} corresponds to the sensing radius $\rho$, and \texttt{expr} is the integrand in Eq.~\ref{eq:nd}. It is the availability of these operators that enables COMSOL Multiphysics\textsuperscript{\textregistered} to solve PIDEs.

In 1D, the non-local interaction term is implemented as
\begin{equation}
I(c,\xi) =
\mathtt{intop} \big(\mathtt{flc2hs}(\rho - |\xi - \xi'|, \delta)\,f\big),
\end{equation}
with a small smoothing value $\delta = 0.001$ used for numerical stability, where
\begin{equation}
f = \frac{c(\xi,\tau)(\xi - \xi')}{|\xi - \xi'| + \varepsilon}.
\end{equation}
The small value $\varepsilon$ is added to the denominator to avoid division by zero at $r = 0$.

In 2D, it is implemented as
\begin{equation}
\vec{I}(c,\vec{\xi}) = 
\begin{bmatrix}
\texttt{diskint}(\rho, f_1, m) \\
\texttt{diskint}(\rho, f_2, m)
\end{bmatrix},
\end{equation}
where
\begin{equation}
f_i = \frac{c(\vec{\xi}, \tau)(\xi_i - \xi'_i)}{r + \varepsilon}.
\end{equation}

The distance $r$ is defined as
\begin{equation}
r = \sqrt{(\xi_1 - \xi'_1)^2 + (\xi_2 - \xi'_2)^2}.
\end{equation}

Here, $(\xi_1, \xi_2)$ are the spatial coordinates of the quadrature points within the disk centered at $(\xi'_1, \xi'_2)$. These coordinates can be accessed using COMSOL's \texttt{dest()} function. For example, the expression $\xi_1-\xi'_1$ is implemented as \texttt{x-dest(x)}.

In 3D, the interaction term is similarly defined using the \texttt{ballint} function:
\begin{equation}
\vec{I}(c,\vec{\xi}) = 
\begin{bmatrix}
\texttt{ballint}(\rho, f_1, m) \\
\texttt{ballint}(\rho, f_2, m) \\
\texttt{ballint}(\rho, f_3, m)
\end{bmatrix},
\end{equation}
where
\begin{equation}
f_i = \frac{c(\vec{\xi}, \tau)(\xi_i - \xi'_i)}{r + \varepsilon}
\end{equation}
with the Euclidean distance in 3D given by
\begin{equation}
r = \sqrt{(\xi_1 - \xi'_1)^2 + (\xi_2 - \xi'_2)^2 + (\xi_3 - \xi'_3)^2}.
\end{equation}

For all simulations shown in the following (except where $m$ is explicitly varied), we use $m = 10$ and $\varepsilon = 10^{-5}$.

\subsection*{Periodic boundary conditions}

\begin{figure}
    \centering
    \includegraphics[width=\linewidth]{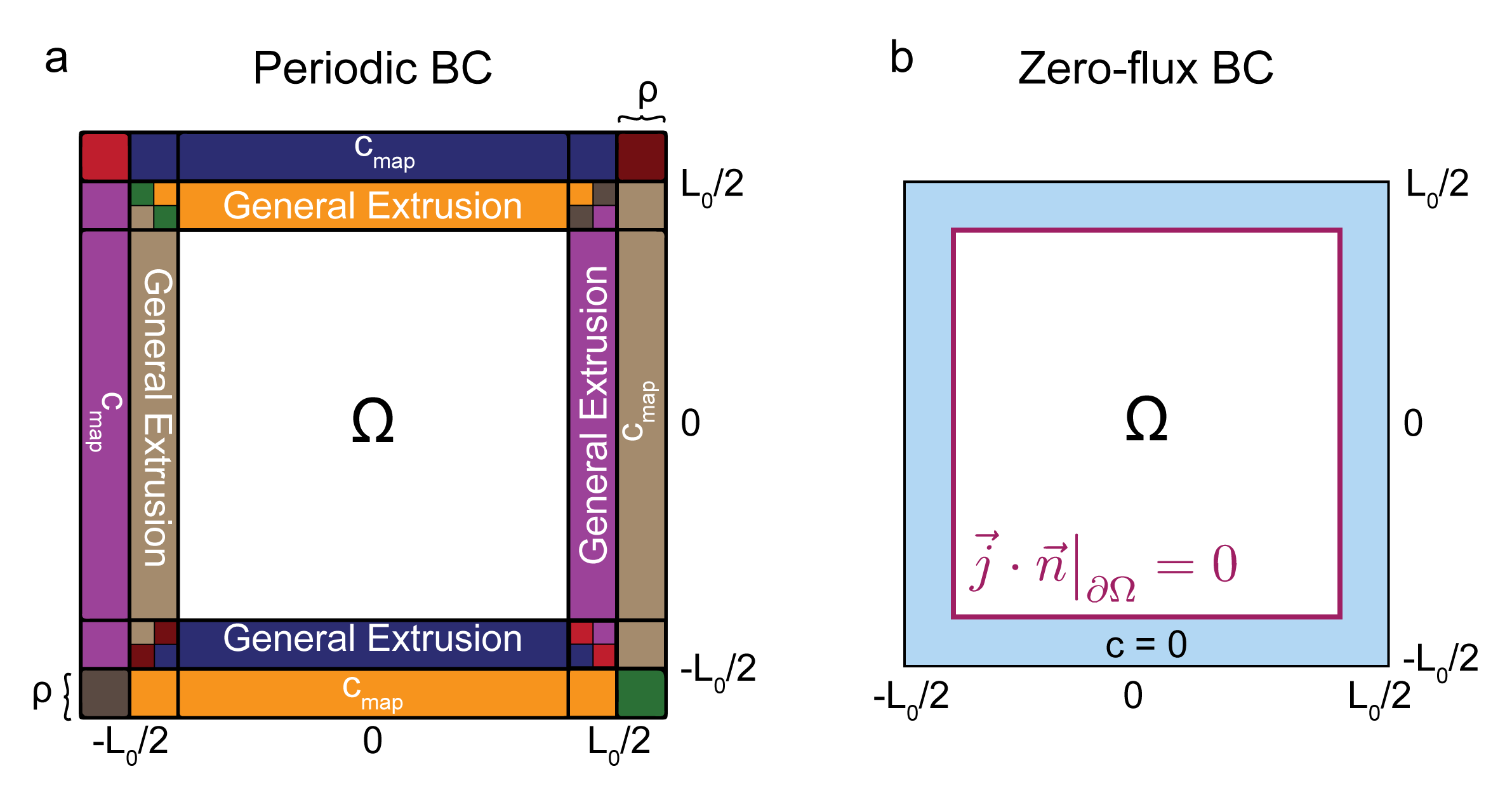}
    \caption{
    \textbf{Illustration of boundary conditions for a 2D domain.}
    \textbf{a} Geometric coordinate mapping for periodic boundary conditions.
    \textbf{b} Zero-flux boundary conditions only need a buffer zone around the simulation domain $\Omega$. 
    }
    \label{fig:bc}
\end{figure}

\begin{figure*}
    \centering
    \includegraphics[width=0.8\textwidth]{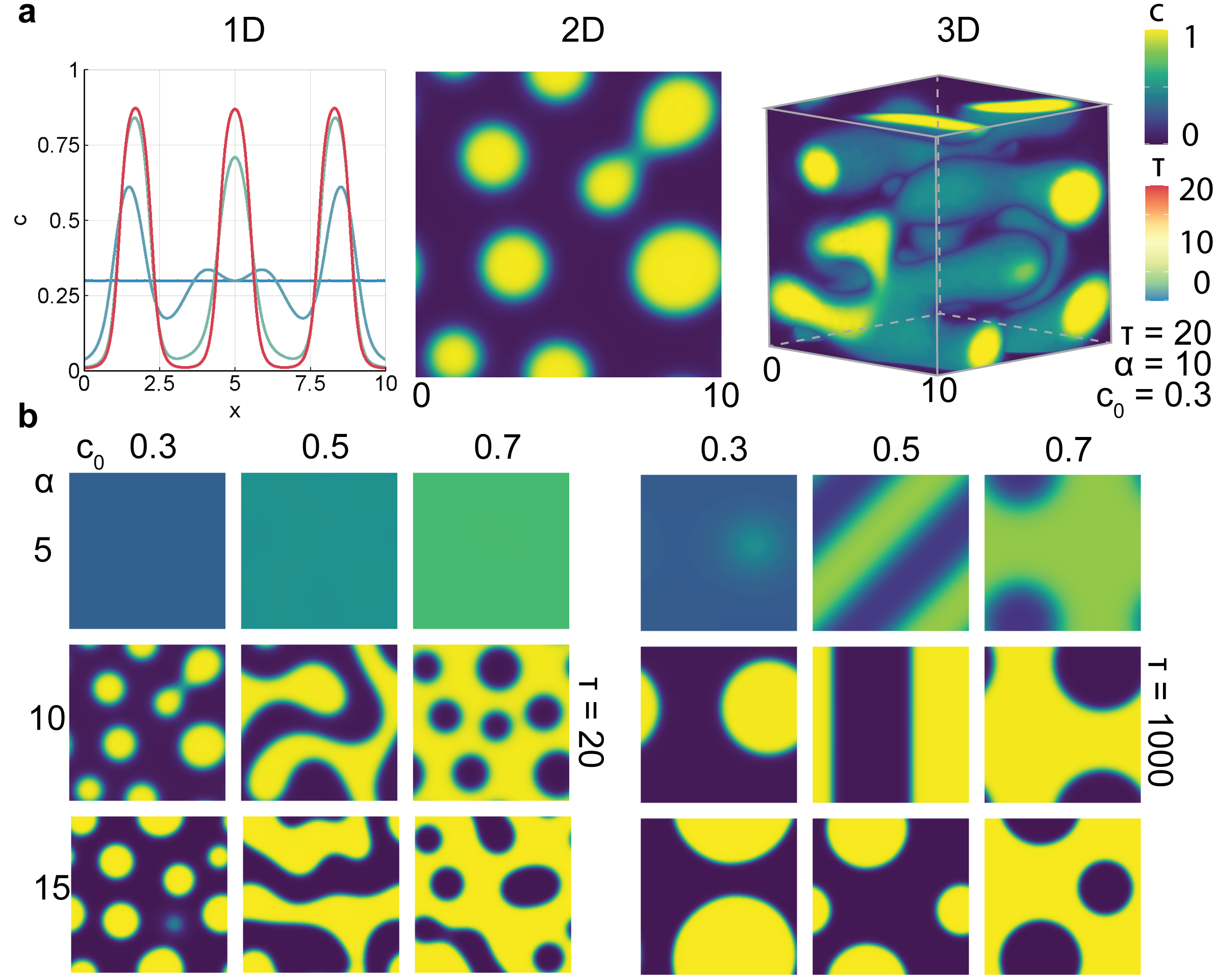}
    \caption{
    \textbf{Summary of the patterning space.}
    \textbf{a} Emergent cell density patterns in one-, two-, and three-dimensional domains with periodic boundary conditions. 
    \textbf{b} Variation of the initial concentration $c_0$ and migratory strength $\alpha$ generates a wide range of patterns. 
    Over time, these patterns coarsen via Ostwald ripening, yielding fewer, larger, and more stable structures.
    }
    \label{fig:qualitative}
\end{figure*}

In the above implementation, it is necessary to ensure that the integral remains well-defined near the domain boundary, where the integration region extends beyond the domain. This was achieved with a coordinate mapping. We illustrate the procedure here in 2D (Fig.~\ref{fig:bc}). For a quadratic domain centered at $(0,0)$, the primary domain (\texttt{Sq1}) of side length $L_0$ is embedded into a concentric auxiliary quadratic domain (\texttt{Sq2}) of side length $L_0 + 2\rho$. For periodic boundary conditions, the positions from the outer square are then mapped to the inner square with the \texttt{General extrusion} operator, using the following rules in 2D:
\begin{equation}
\begin{cases}
x \to x + L_0 & \text{if } x \in [-L_0/2 - \rho,\; -L_0/2] \\
x \to x - L_0 & \text{if } x \in [+L_0/2,\; +L_0/2 + \rho] \\
y \to y + L_0 & \text{if } y \in [-L_0/2 - \rho,\; -L_0/2] \\
y \to y - L_0 & \text{if } y \in [+L_0/2,\; +L_0/2 + \rho]
\end{cases}
\end{equation}
For example, to implement the mapping  
\begin{equation}
y \;\to\; y + L_0 
\quad \text{for} \quad y \in \big[-\!L_0/2 - \rho,\; -L_0/2\big] 
\end{equation}
(orange domain), one defines a new \texttt{General extrusion} operator. In this case, the three upper domains are selected, with the source region ranging from
\begin{equation}
(x,\;y) = (-L_0/2,\; +L_0/2-\rho) 
\quad \text{to} \quad 
(+L_0/2,\; +L_0/2).
\end{equation}
The destination map is then specified by setting the \texttt{x-expression} to $x$ as we only want to shift the values in $y$ direction,
and the \texttt{y-expression} to $y + L_0$, which maps the selected region onto the orange domain labeled $c_\text{map}$, ranging from
\begin{equation}
(x,\;y) = (-L_0/2 ,\; -L_0/2 - \rho) 
\quad \text{to} \quad 
(+L_0/2,\; -L_0/2).
\end{equation}
This procedure is then repeated for all four edges and four corners, totaling in eight mappings in 2D.

For 3D, the squares are replaced by cubes centered at $(0, 0, 0)$ and the $z$-component is added to the mapping in the form of:
\begin{equation}
\begin{cases}
    z \to z + L_0 & \text{if } z \in [-L_0/2 - \rho,\; - L_0/2] \\
    z \to z - L_0 & \text{if } z \in [+L_0/2,\; +L_0/2 + \rho]
\end{cases}
\end{equation}
The total number of coordinate mappings is then 26 (6 faces, 12 edges, 8 corners).
 
\subsection*{Zero-flux boundary conditions}

Zero-flux boundary conditions present a similar challenge to periodic boundaries, as the non-local integral must also be defined at the domain boundaries. Unlike periodic boundaries, however, this issue can easily be resolved by setting the solution in the added outer rim (\texttt{Sq2}) to zero, which ensures that the integral is evaluated consistently across the entire physical domain, including its edges. On the domain boundary $\partial \Omega$, the flux is then simply constrained by a zero-flux (Neumann) condition:
\begin{equation}
\vec{j} \cdot \vec{n}\big|_{\partial \Omega} = 0.
\end{equation}

\subsection*{Initial Conditions}

The initial condition of the system is a spatially uniform concentration $c_0$, perturbed by random biological density fluctuations of amplitude $\sigma$ to break the symmetry. For simplicity, these fluctuations are sampled from a uniform distribution with vanishing mean:
\begin{equation}
c(\vec{\xi},0) = c_0 \left(1 + \sigma X(\vec{\xi})\right),\quad X \sim \mathcal{U}(-0.5,\ 0.5).
\end{equation}

\subsection*{Solver settings}

We used an implicit time stepping method called the backward differentiation formula (BDF) with the order limited from one to five. The relative error tolerance was set to $10^{-4}$ for all 2D simulations and $10^{-3}$ for the more expensive 3D simulations. The computational domain was discretized into triangular or tetrahedral elements, and quadratic Langrangian shape functions were used. The maximum mesh element size was always expressed in terms of $\rho$ and was, unless otherwise noted, $\rho/2$.

\section*{Results}

The simulations can be carried out in one-, two, and three dimensional domains (Fig.~\ref{fig:qualitative}a). Spatial and temporal patterning can be modulated to transition from spots to labyrinth shapes to inverse spots (Fig.~\ref{fig:qualitative}b). The key parameters to modulate here are the migration activity $\alpha$, with greater values leading to accelerate pattern formation (or enabling patterning in the first place), and the initial density $c_0$, modulating the patterning phenotype.
Not all parameter combinations lead to stable simulations, for example $c_0$ values close to 0.5, large $
\alpha$ and $\rho$ lead to numerical instabilities. A detailed stability analysis and further quantitative insights are provided elsewhere \cite{yu2025directed}.
The examples were implemented using the parameters listed in Table~\ref{tab:params} unless otherwise noted.

\begin{table}
\centering
\caption{Default simulation parameters.}
\label{tab:params}
\begin{tabular}{lll}
\toprule
Parameter & Value     & Description                  \\ \midrule
$\alpha$         & 10        & Interaction strength                  \\
$\rho$      & 1        & Sensing radius                   \\
$L_0$          & $10 \rho$        & (Initial) domain side length \\
$c_0$     & 0.3       & Initial cell density         \\
$\sigma$     & 0.01       & Initial noise level         \\
$\tau_\mathrm{end}$      & 20        & Final time point                    \\
$v$        & 0        & Tissue growth rate\\
$h$        & $\rho/2$ & Mesh element side length                    \\
$m$          & 10        & Number of integration points \\
\bottomrule
\end{tabular}
\end{table}

\begin{figure}
    \centering
    \includegraphics[width=\columnwidth]{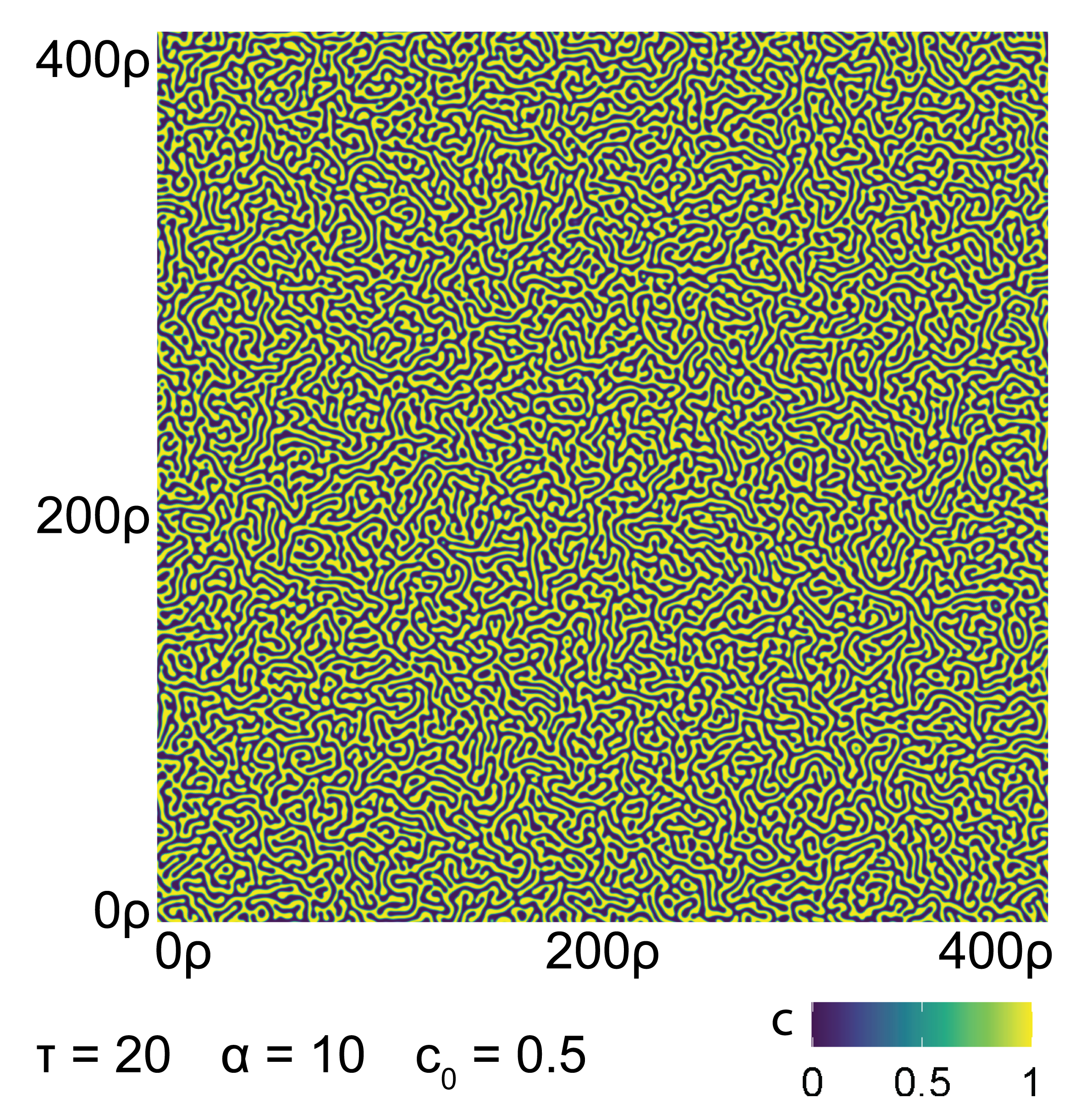}
    \caption{\textbf{Exemplary simulation on a large domain.} 
    The simulation was performed on a square domain of side length $400\rho$, with initial concentration $c_0 = 0.5$, migratory strength $\alpha = 10$, and integrated until $\tau_\mathrm{end} = 20$. The computational mesh was composed of 1,832,722 elements and 917,970 nodes, resulting in 3,630,085 degrees of freedom. The simulation ran for 16\,h and 41\,min on 8 cores of two Intel Xeon Gold 6244 CPUs (3.60\,GHz).
}
    \label{fig:large}
\end{figure}

\begin{figure*}
    \centering
    \includegraphics[width=0.9\textwidth]{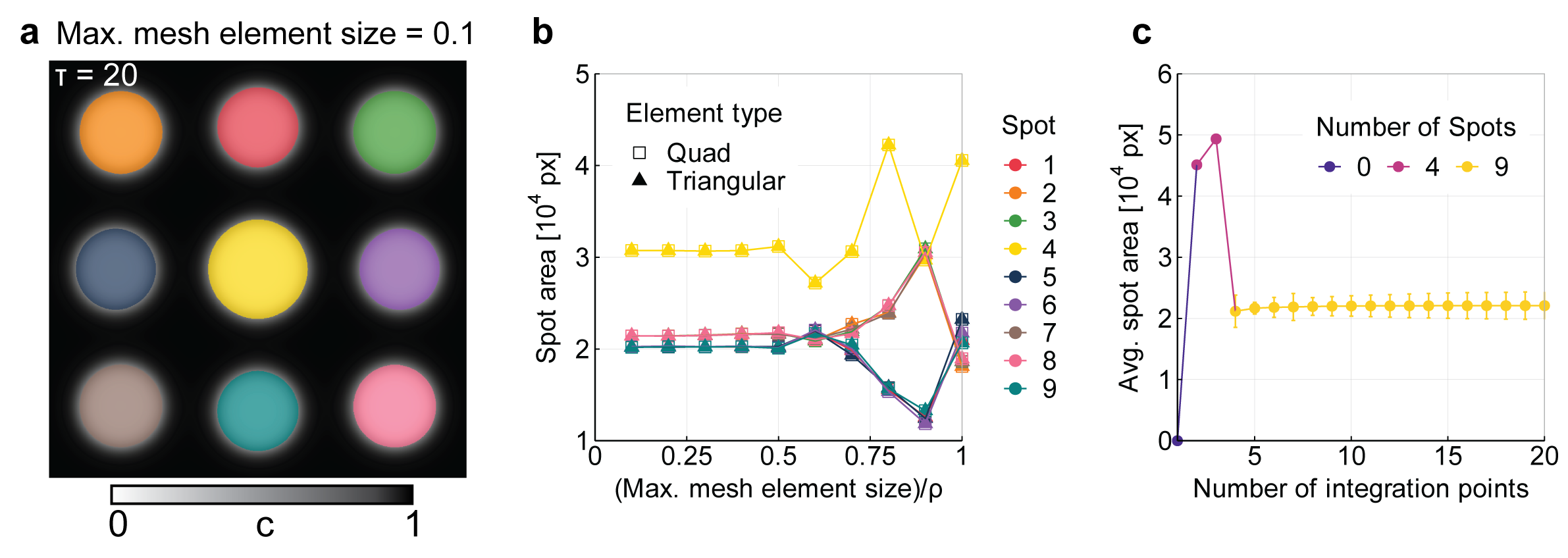}
    \caption{\textbf{Impact of mesh element size and number of integration points.}
    \textbf{a} Example of segmented spots for a mesh with maximum element size of $0.1\rho$.
    \textbf{b} Scaling of spot area with mesh resolution. 
    \textbf{c} Impact of the number of integration points for the integral on the pattern.
    For all simulations, zero flux boundary conditions, $\alpha = 10$ and $c_0 = 0.3$ were used. Spots were analyzed at $\tau = 20$.
    }
    \label{fig:mesh_size}
\end{figure*}

The implementation is scalable to large domain sizes, as the effective characteristic length scale of the pattern depends mainly on the sensing radius. The implementation is therefore suitable for investigating a broad range of biologically relevant processes. We demonstrate this with a large-scale simulation on a domain of size $L_0=400\rho$ (Fig.~\ref{fig:large}). In biological systems, sensing radii---for instance mediated by cellular protrusions such as cytonemes---typically range between $10$ and $150$ \textmu{}m \cite{daly2022regulatory}. This corresponds to tissue sizes of $0.16$--$36$ cm\textsuperscript{2}, placing the simulations well within biologically relevant scales.

To investigate the impact of the mesh size and mesh element type, we computed a set of simulations using a fixed parameter set and varied the mesh edge size from $0.1\rho$ to $1\rho$. For better comparability, all simulations were done with zero-flux boundary conditions. We segmented the resulting cell density image using basic thresholding (Fig.~\ref{fig:mesh_size}a). For all tested element sizes, nine spots appeared. Comparing the spot areas for simulations with varying maximum mesh element size, we found a consistent, stable behavior up to a mesh element size of $0.5\rho$ (Fig.~\ref{fig:mesh_size}b). For elements larger than that, the spot size becomes unstable and starts to vary. The difference between quad (squares) and triangular (triangles) mesh element types is negligible.

For PIDEs, the numerical discretization of the integral term plays a critical role. In particular, the number of integration points can strongly influence the simulation outcome. A greater number of points increases integration accuracy but also incurs greater computational cost, whereas too few points can lead to artifacts in the solution. Specifically, we observe instability for fewer than four integration points in 2D, yielding four large spots instead of the expected nine (Fig.~\ref{fig:mesh_size}c). In contrast, simulations with five or more integration points produce stable and consistent patterns. This can be understood by the requirement that cells need to probe their environment in all $2n$ directions in $n$-dimensional space for the patterning model to work.

\begin{figure*}
    \centering
    \includegraphics[width=0.85\textwidth]{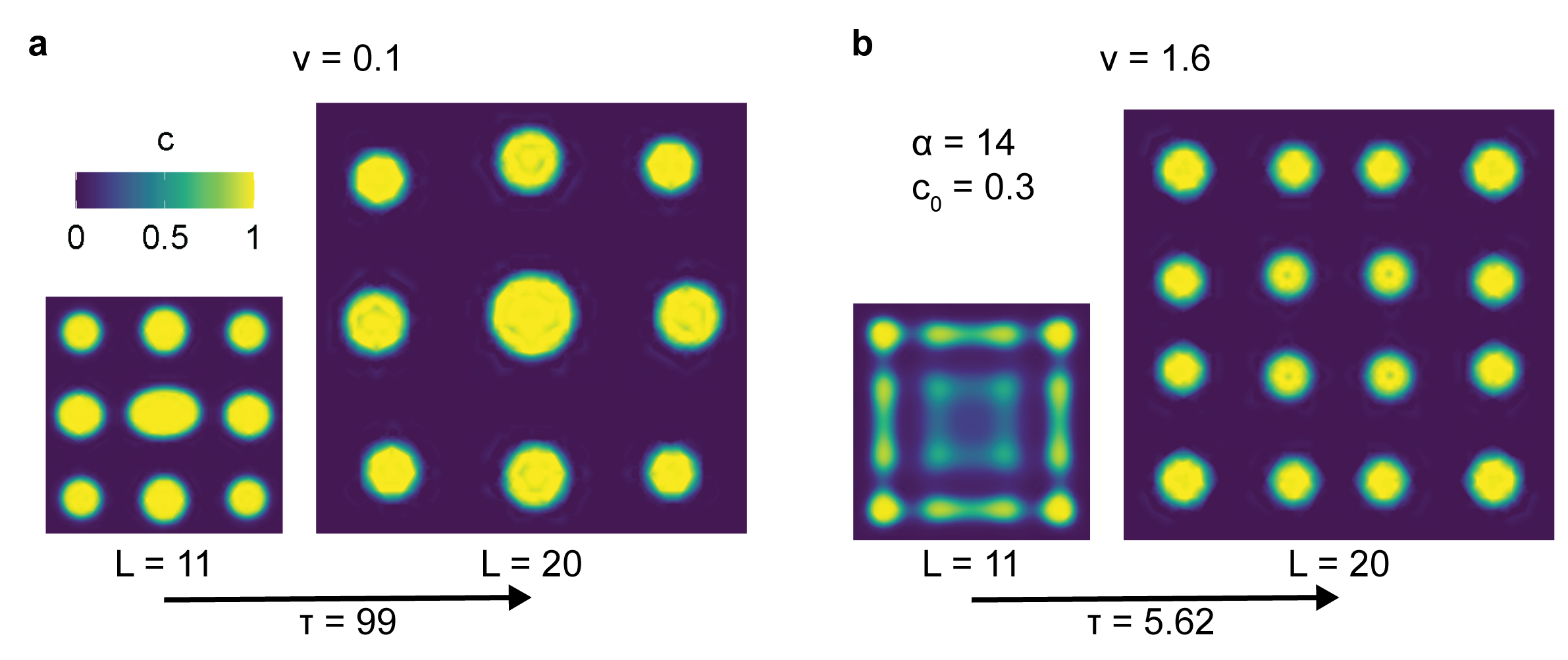}
    \caption{\textbf{Impact of growth speed on patterning.}
    Slower growth (\textbf{a}) leads to fewer, more spaced out spots than faster growth (\textbf{b}) on domains of the same size. 
    Slower growth was simulated with a growth speed of $v = 0.1$ for $\tau_\mathrm{end}=100$, the faster growth with $v = 1.6$ for $\tau_\mathrm{end}=6.25$.
    }
    \label{fig:growth}
\end{figure*}

During development, most pattern formation processes, including directed cell migration, typically occur concurrently with tissue growth. By incorporating domain growth into the model, we demonstrate how the growth rate modulates the emergent patterning phenotype (Fig.~\ref{fig:growth}). At a slower growth rate ($v=0.1$), the system produces nine more widely spaced spots, whereas a faster growth rate ($v=1.6$) yields $16$ smaller spots under identical parameter settings ($\alpha = 14$, $c_0 = 0.3$). This shift arises because rapid domain expansion effectively rescales the intrinsic patterning wavelength, permitting the transient nucleation of additional spots. The results are robust across repeated simulations, highlighting growth rate as a key control parameter in pattern selection \cite{yu2025directed}.

\section*{Conclusions}

In conclusion, we present a generalizable PIDE-based model formulation for directed cell migration, implemented within the general-purpose finite-element framework COMSOL Multiphysics\textsuperscript{\textregistered}. This approach provides a powerful tool to investigate a broad spectrum of cell motility–driven processes in biology. Its applicability across one-, two-, and three-dimensional domains ensures versatility for diverse modeling scenarios, while its computational efficiency enables simulations at tissue scale, including realistic conditions such as dynamically growing domains. Looking ahead, this framework opens opportunities for the integration of additional biophysical mechanisms, extension to heterotypic cellular interactions, and systematic exploration of how tissue-scale dynamics emerge from processes at the cellular scale. In particular, it provides a promising avenue to study biologically relevant processes such as morphogen-mediated tissue patterning, wound healing, and cancer invasion.

\FloatBarrier

\section*{Acknowledgements}
The authors thank Walter Frei and Sven Friedel from COMSOL Inc.\ for technical assistance with the numerical implementation, Vincent Zanetta and Barbara Walkowiak for preliminary model assessment, as well as the CoBi group at ETH Z\"{u}rich, Kevin J.\ Painter and Jos\'{e} A.\ Carrillo for discussions.


\begin{thebibliography}{16}
\providecommand{\natexlab}[1]{#1}
\providecommand{\url}[1]{\texttt{#1}}
\expandafter\ifx\csname urlstyle\endcsname\relax
  \providecommand{\doi}[1]{doi: #1}\else
  \providecommand{\doi}{doi: \begingroup \urlstyle{rm}\Url}\fi

\bibitem[Yu et~al.(2025)Yu, Mederacke, Vetter, and Iber]{yu2025directed}
C.~Yu, M.~Mederacke, R.~Vetter, and D.~Iber.
\newblock Directed cell migration is a versatile mechanism for rapid
  developmental pattern formation.
\newblock \emph{BioRxiv}, 2025.
\newblock \doi{10.1101/2025.07.24.666657}.

\bibitem[Germann et~al.(2011)Germann, Menshykau, Tanaka, and
  Iber]{Germann:2011}
P.~Germann, D.~Menshykau, S.~Tanaka, and D.~Iber.
\newblock {Simulating Organogenesis in COMSOL}.
\newblock In \emph{{Proceedings of the 2011 COMSOL Conference in Stuttgart}},
  2011.

\bibitem[Menshykau and Iber(2012)]{Menshykau:2012}
D.~Menshykau and D.~Iber.
\newblock {Simulation Organogenesis in COMSOL: Deforming and Interacting
  Domains}.
\newblock In \emph{{Proceedings of the 2012 COMSOL Conference in Milan}}, 2012.

\bibitem[Menshykau et~al.(2013)Menshykau, Shrivastsan, Germann, Lemereux, and
  Iber]{Menshykau:2013}
D.~Menshykau, A.~Shrivastsan, P.~Germann, L.~Lemereux, and D.~Iber.
\newblock {Simulating Organogenesis in COMSOL Multiphysics: Parameter
  Optimization for PDE-based Models}.
\newblock In \emph{{Proceedings of the 2013 COMSOL Conference in Rotterdam}},
  2013.

\bibitem[Vollmer et~al.(2013)Vollmer, Menshykau, and Iber]{Vollmer:2013}
J.~Vollmer, D.~Menshykau, and D.~Iber.
\newblock {Simulating Organogenesis in COMSOL Multiphysics: Cell-based
  Signaling Models}.
\newblock In \emph{{Proceedings of the 2013 COMSOL Conference in Rotterdam}},
  2013.

\bibitem[Karimaddini et~al.(2014)Karimaddini, \"{U}nal, Menshykau, and
  Iber]{Karimaddini:2014}
Z.~Karimaddini, E.~\"{U}nal, D.~Menshykau, and D.~Iber.
\newblock {Simulating Organogenesis in COMSOL Multiphysics: Image-Based
  Modeling}.
\newblock In \emph{{Proceedings of the 2014 COMSOL Conference in Cambridge}},
  2014.

\bibitem[Wittwer et~al.(2016)Wittwer, Croce, Aland, and Iber]{Wittwer:2016}
L.~D. Wittwer, R.~Croce, S.~Aland, and D.~Iber.
\newblock {Simulating Organogenesis with COMSOL Multiphysics software:
  Phase-Field Based Simulations of Embryonic Lung Branching Morphogenesis}.
\newblock In \emph{{Proceedings of the 2016 COMSOL Conference in Munich}},
  2016.

\bibitem[Wittwer et~al.(2017)Wittwer, Peters, Aland, and Iber]{Wittwer:2017}
L.~D. Wittwer, M.~D. Peters, S.~Aland, and D.~Iber.
\newblock {Simulating Organogenesis in COMSOL Multiphysics: Comparison of
  Methods for Simulating Branching Morphogenesis}.
\newblock In \emph{{Proceedings of the 2017 COMSOL Conference in Rotterdam}},
  2017.

\bibitem[Peters and Iber(2017)]{Peters:2017}
M.~D. Peters and D.~Iber.
\newblock {Simulating Organogenesis in COMSOL Multiphysics: Tissue Mechanics
  during Organ Growth}.
\newblock In \emph{{Proceedings of the 2017 COMSOL Conference in Rotterdam}},
  2017.

\bibitem[Conrad et~al.(2021)Conrad, Runser, Fernando~Gomez, Lang, Dumond,
  Sapala, Schaumann, Michos, Vetter, and Iber]{Conrad:2021}
L.~Conrad, S.~V.~M. Runser, H.~Fernando~Gomez, C.~M. Lang, M.~S. Dumond,
  A.~Sapala, L.~Schaumann, O.~Michos, R.~Vetter, and D.~Iber.
\newblock The biomechanical basis of biased epithelial tube elongation in lung
  and kidney development.
\newblock \emph{Development}, 148\penalty0 (9), 2021.
\newblock \doi{10.1242/dev.194209}.

\bibitem[Kurics et~al.(2014)Kurics, Menshykau, and Iber]{Kurics:2014}
T.~Kurics, D.~Menshykau, and D.~Iber.
\newblock Feedback, receptor clustering, and receptor restriction to single
  cells yield large turing spaces for ligand-receptor-based turing models.
\newblock \emph{Phys Rev E Stat Nonlin Soft Matter Phys}, 90\penalty0
  (2):\penalty0 022716, 2014.
\newblock \doi{10.1103/PhysRevE.90.022716}.

\bibitem[Carrillo et~al.(2019)Carrillo, Murakawa, Sato, Togashi, and
  Trush]{carrillo2019population}
J.~A. Carrillo, H.~Murakawa, M.~Sato, H.~Togashi, and O.~Trush.
\newblock A population dynamics model of cell-cell adhesion incorporating
  population pressure and density saturation.
\newblock \emph{J. Theor. Biology}, 474:\penalty0 14--24, 2019.
\newblock \doi{10.1016/j.jtbi.2019.04.023}.

\bibitem[Falc et~al.(2023)Falc, Baker, and Carrillo]{falco2023local}
C.~Falc, R.~Baker, and J.~Carrillo.
\newblock A local continuum model of cell-cell adhesion.
\newblock \emph{SIAM J. Appl. Math.}, 84:\penalty0 S17--S42, 2023.
\newblock \doi{10.1137/22M1506079}.

\bibitem[Armstrong et~al.(2006)Armstrong, Painter, and
  Sherratt]{armstrong2006continuum}
N.~J. Armstrong, K.~J Painter, and J.~A. Sherratt.
\newblock A continuum approach to modelling cell-cell adhesion.
\newblock \emph{J. Theor. Biology}, 243:\penalty0 98--113, 2006.
\newblock \doi{10.1016/j.jtbi.2006.05.030}.

\bibitem[Carrillo et~al.(2025)Carrillo, Murakawa, Sato, and
  Wang]{carrillo2025new}
J.~A. Carrillo, H.~Murakawa, M.~Sato, and M.~Wang.
\newblock A new paradigm considering multicellular adhesion, repulsion and
  attraction represent diverse cellular tile patterns.
\newblock \emph{PLOS Comput. Biol.}, 21:\penalty0 e1011909, 2025.
\newblock \doi{10.1371/journal.pcbi.1011909}.

\bibitem[Daly et~al.(2022)Daly, Hall, and Ogden]{daly2022regulatory}
C.~A. Daly, E.~T. Hall, and S.~K. Ogden.
\newblock Regulatory mechanisms of cytoneme-based morphogen transport.
\newblock \emph{Cell. Mol. Life Sci.}, 79:\penalty0 119, 2022.
\newblock \doi{10.1007/s00018-022-04148-x}.

\end{thebibliography}
\end{document}